# Creasing of an everted elastomer tube

Xudong Liang, Feiyu Tao, Shengqiang Cai*

*Department of Mechanical and Aerospace Engineering, University of California, San Diego,*

*La Jolla, CA 92093, USA*

**Abstract**

A cylindrical elastomer tube can stay in an everted state without any external forces. If the thickness of the tube is small, the everted tube, except for the regions close to the free ends of the tube, maintains cylindrical shape; if the thickness is larger than a critical value, the cross-section of the everted tube becomes noncircular, which is caused by mechanical instability. Although eversion-induced instability in an elastomer tube has been reported several decades before, a satisfying explanation of the phenomenon is still unavailable. In all previous studies, linear analyses have been adopted to predict the critical thickness of the tube for the eversion-induced instability. The discrepancy between the prediction and experiment is significant. In this letter, based on the experiments and theoretical analyses, we show that crease formation on the inner surface of an everted tube is the mechanical instability mode, which cannot be captured by linear stability analyses. Instead, a combination of energetic analyses and numerical simulations of finite deformation in an everted tube enables us to correctly predict both critical tube thickness for the onset of creases and profile of the noncircular cross-section of an everted tube with large thickness.

* Correspondent author

E-mail: s3cai@ucsd.edu (Shengqiang Cai)

Turning a structure inside out, often called eversion, is ubiquitous in nature and frequently used in different fabrication processes. For example, during the development of *Volvox* embryo, a spherical monolayer cell sheet turns itself inside out to achieve its adult configuration [1]; the eversion of jellyfish, known as 'Jellyfish syndrome', is a mechanism to protect itself from environmental changes [2,3]; stent eversion has been used to construct an autologous heart valve, which is known as the stent-biovalve [4]. Due to its profound influence in nature and in various engineering applications, eversion of 3D objects has been a topic of significant interest for a large group of researchers for several decades [1, 5-13].

Finite deformation is often involved in the eversion of various structures [12]. Deformation of an everted elastomeric tube has been one of the most classical finite deformation problems since Rivilin [12] first proposed it. Varga [8] has shown that in experiments, except for the regions close to the two ends of the tube, most part of the tube after eversion is very close to be the cylindrical shape. Chadwick and Haddon have investigated the conditions for the existence and uniqueness of the solutions associated with cylindrical tube eversion [9]. Eriksen [10] and Antman [11] further extended the results to the eversion of spherical shells.

However, an experimental phenomenon associated with the tube eversion, first described by Truesdell [13], has not been well explained for several decades. Truesdell discovered in his experiment that the cross-section of an everted cylindrical tube became noncircular when the tube thickness is large enough. To explain the observation quantitatively, both linear stability analyses [5, 6] and weakly nonlinear analyses of tube eversion [7] have been conducted by different researchers. These analyses have shown that when the tube thickness is larger than a critical value, axisymmetric deformation of an everted tube is not stable any more and wrinkles may appear on the inner surface of the tube after eversion. Nevertheless, as pointed out in many of these articles [5, 6], the critical thickness of the tube predicted by all the previous analyses is significantly larger than the thickness of the tube used in Truesdell's experiment. The well-known discrepancy between the predictions and experiments has not been resolved until now.

In the current study, we fabricate cylindrical elastomer tubes with various sizes using homemade mold. The material of the tube is a silicone rubber purchased from the company Smooth-on (USA). We, for the first time, experimentally demonstrate that when a cylindrical tube is everted as shown in Fig. 1, multiple creases may form on its inner surface, which has been recently identified as a distinct surface instability mode from wrinkles [14-16]. Previous studies have also shown that due to the highly nonlinear deformation associated with crease formation, the critical condition for crease initiation cannot be predicted using linear or weakly nonlinear analyses [15]. In this article, through energetic analyses and linear perturbation calculations, we show that creasing instability, instead of wrinkling, should be responsible for the mechanical instability associated with tube eversion.

To study the instability of an everted tube, we first calculate the axisymmetric deformation field of an everted cylindrical tube as shown in Fig. 2a. We use the polar coordinate $\mathbf{X}=(R,\Theta,Z)$ to describe the undeformed configuration of the tube and $\mathbf{x}=(r,\theta,z)$ for its deformed configuration. As shown in Fig. 2a, the inner and outer radius of the tube in the undeformed state are $A$ and $B$, and change to $a$ and $b$ in the everted state. Previous studies have shown that by neglecting the edge effect, the deformation of an everted tube can be simply described by the following field [5],

$$r = r(R), \Theta = \theta, z = \lambda Z. \tag{1}$$

The stretch in the radial direction is $\lambda_r = -dr/dR$, in the hoop direction is $\lambda_\theta = r/R$, and in the axial direction is $\lambda_z = \lambda$.

With the assumption of axisymmetric deformation, force balance equation for an everted tube is,

$$\frac{d\sigma_{rr}}{dr} + \frac{\sigma_{rr} - \sigma_{\theta\theta}}{r} = 0. \tag{2}$$

where $\sigma_{rr}$ and $\sigma_{\theta\theta}$ are Cauchy stress in the radial direction and hoop direction, respectively. The boundary condition for the everted tube is $\sigma_{rr} = 0$ on the inner surface: $r = a$, and the outer surface: $r = b$. Following Rivilin [12], instead of requiring the normal stress to be zero on the ends of the tube, we relax the boundary condition to enforce the resultant force applied on the ends of the tube to be zero, namely, $\int_a^b r\sigma_{zz} dr = 0$. As shown in the early experiments done by Varga [8], the edge effect in an everted tube only exists for the region around the distance of the thickness of the tube away from its free ends.

For simplicity, we assume that the material of the tube can be described by incompressible Neo-Hookean model [17]. Therefore, we have the following stress-stretch relationship for the material: $\sigma_{rr} = \mu\lambda_r^2 - p$, $\sigma_{\theta\theta} = \mu\lambda_\theta^2 - p$ and $\sigma_{zz} = \mu\lambda_z^2 - p$, where $\mu$ is small-deformation shear modulus of the elastomer and $p$ is hydrostatic pressure. The material is assumed to incompressible, so we have $\lambda_r \lambda_\theta \lambda_z = 1$.

The equilibrium solution of an everted tube can be numerically solved using shooting method and the details are discussed in the *Appendix A*. Because both the stress boundary conditions and stress balance equations are homogenous, the equations apparently have a trivial solution corresponding to the state without deformation. A nontrivial solution corresponds to the everted state of the tube with finite deformation. In Fig. 2b, we plot the normalized inner radius *a/B*, the normalized outer radius *b/B* and axial stretch $\lambda$ of the everted tube as functions of the normalized thickness of the tube *H/B* (*H=B-A*). After eversion, the axial stretch $\lambda$ of the tube is always larger than one. When the thickness of the tube *H* approaches zero, the axial stretch of the everted tube is close to one. As the tube thickness *H* increases, the axial stretch of the everted tube also increases. Outer radius of the everted tube is always smaller than that of the tube in the undeformed state. The comparison between the experimental measurements and theoretical predictions of the inner and outer radius of the everted tube for a range of thickness *H* has been shown in Fig. 2b.

The corresponding stretch and stress field along the radius of an everted tube with the thickness *H=0.5B* are plotted in Fig. S1. In the radial direction, the stretch $\lambda_r$ decreases from the value larger than one in the outer surface to the value smaller than one on the inner surface. The radial stress $\sigma_{rr}$ between the inner and outer surface of the tube is always compressive. In the hoop direction, both the stretch $\lambda_\theta$ and stress $\sigma_{\theta\theta}$ are compressive on the inner surface of the everted tube, which may result in surface instability that is shown in Fig. 1b.

As discussed before, wrinkling and creasing are two distinct but commonly observed surface instability modes in soft materials [14-16, 18]. Wrinkles are usually characterized by a smooth undulation while creases are characterized by singular regions of self-contact. Linear perturbation analysis has been usually adopted to obtain the critical condition of wrinkling instability [19], while a combination of numerical simulation and energetic analyses has been used to study the crease formation [15]. In the following, we calculate the critical conditions for both wrinkling and creasing instability of a tube in everted state, respectively.

With the assumption of axisymmetric deformation of an everted tube, we have calculated the deformation field $r = r(R)$ of an everted tube as shown in *Appendix A*. To conduct linear stability analysis, we perturb the axisymmetric deformation field $r = r(R)$ by a displacement field in both radial direction: $u_r(r,\theta)$ and hoop direction: $u_\theta(r,\theta)$, which further result in the perturbations of both stretch field and stress field. Assuming the perturbations are infinitesimal, we only keep the linear terms of displacement perturbations in the governing equations. Consequently, we obtain an eigenvalue problem (B10~B12 in *Appendix B*). The characteristic equation determines the critical condition, namely the tube thickness *H*, for the onset of wrinkles, and the associated eigenvectors describe the mode of wrinkling. Detailed formulation of linear stability analyses can be found in *Appendix B*. In Fig. S3 we plot the critical tube thickness *H* for the onset of wrinkles with different wavenumbers. From the calculations, we can conclude that the critical tube thickness for the wrinkling instability associated with eversion is $H_{crit}$=0.58*B*. Similar results have also been reported by Haughton et al [5, 6].

Next, we aim at obtaining the critical condition for creasing instability on the inner surface of an everted tube. As shown in our previous studies [16], the crease initiation is autonomous, so the critical condition of creasing can be determined by the local strain field. Based on numerical simulation and energetic analyses, Hong et al [15] obtained the critical condition for the initiation of a crease on a free surface: $\lambda_1 / \lambda_2 = 2.4$, where $\lambda_1$ and $\lambda_2$ are the two principal stretches on the surface, and they correspond to $\lambda_r$ and $\lambda_\theta$ in cylindrical coordinate adopted in current tube eversion problem. Comparing the equilibrium solution in Fig. S2 and the critical condition for crease initiation, we can obtain that the critical tube thickness for creasing instability is $H_{crit=}$0.435*B*, which is dramatically smaller than the critical thickness for wrinkling instability predicted based on linear stability analysis. Therefore, creasing is the surface instability mode for an everted tube, which explains why all the previous linear stability analyses overestimated the critical tube thickness. Our statement has also been partially confirmed by the experiment shown in Fig. 1. In Fig. 1a, the inner surface of an everted tube is smooth with the thickness: *H*=0.42*B*; while in Fig. 1b, creases are visible on the inner surface of the everted tube for the thickness: *H*=0.5*B*. Although it is not easy to determine the exact thickness for the onset of crease, the critical thickness of the tube has to fall in the range between 0.42*B* and 0.5*B*.

We next conduct finite element simulations to further verify the critical condition for the crease initiation obtained above and investigate post-creasing phenomenon of an everted tube. To avoid simulating the complex eversion process, we map our previously computed stress field for the axisymmetric deformation of an everted tube into our 2D finite element model. The axial stretch $\lambda$ is kept as a constant in the simulation, so it is a generalized plane strain problem. To simplify the problem, we assume the creases distribute around the inner surface of an everted tube periodically. Therefore, only a sector of the cross-section of a tube is used in the simulation (inset of Fig. 3). By using the sectors with different angles $\theta$, we can simulate the scenario of an everted tube with different number of creases on its inner surfaces. To artificially introduce a crease, we impose a radial displacement at a point on the inner surface of an everted tube as shown in Fig. 3. We then numerically calculate the strain energy of the everted tube as a function of crease depth $d$. To accurately capture the stress/strain field associated with the formation and growth of creases in an everted tube and improve the convergence of the numerical simulation, we adopt mesh-to-mesh mapping technique in the simulation (detailed discussion of the simulation procedure can be found in *Appendix C*). According to our knowledge, the numerical simulation strategy described above has never been reported to model creasing instability in previous studies. In Fig. 3, we plot the strain energy of the everted tube in creased state $U_c$, normalized by the strain energy of the everted tube with asymmetric smooth deformation $U_o$, as a function of the normalized crease depth $d/H$. When the tube thickness $H$ is small, the strain energy of an everted tube increases monotonically with increasing in the crease depth $d$. When the tube thickness $H$ is larger than a critical value, namely $0.435B$, the strain energy of the everted tube decreases first with increasing the crease depth $d$, which reaches a minimum for a certain crease depth. Further increase of crease depth will cause the increase of the strain energy. The computational results indicate that when the tube thickness is larger than the critical thickness, the formation of creases on the inner surface of the everted tube can reduce the strain energy. Therefore, the numerical simulation is consistent with our previous predictions of the critical tube thickness for crease formation.

To further predict the total number of creases on the inner surface of an everted tube, by varying the crease depth and the angle of the sector in our simulation, we can compute the minimal strain energy of the everted tube with different number of creases. As shown in Fig. 4a, when the thickness of the tube $H$ is larger than the critical thickness, an everted tube with finite number of creases on its inner surface minimizes the total strain energy. Based on the computational results shown in Fig. 4a, we further plot the number of creases, which minimize the total strain energy of the everted tube, as a function of the tube thickness. With increase the tube thickness, the number of creases on the inner surface of everted tubes decreases. The cross-section of an everted tube with thickness of $H/B$=0.52 between the predicted result and microCT image are compared in Fig. 5. It clearly shows that not only the number of creases, but also the profile of the inner surface of the everted tube can be well captured by our predictions.

Although our predictions agree well with experimental observations, certain limitations exist in our theory. For example, we assume the distributions of creases are periodic in the inner surface of everted tubes. The assumption may be invalid in certain scenarios. In particular, preexisting defects in the system may predetermine the locations of crease formation. In addition, we adopt Neo-Hookean model to characterize the hyperelasticity of elastomers in our experiments, which may be inaccurate for different elastomers. However, as discussed in the reference [15], the strain involved in the crease formation is finite but modest; Neo-Hookean model should be accurate enough for most elastomers.

In summary, we studied mechanical instability of an everted cylindrical elastomer tube. By comparing the critical conditions of creasing and wrinkling, we show that the formation of creases, instead of wrinkles, is the surface instability mode for an everted tube when its thickness is larger than a critical value. Our studies have successfully resolved a long-lasting discrepancy between the theoretical predictions and experimental observations of the critical thickness for an everted tube being unstable. Based on the minimal strain energy assumption, we have also successfully predicted the number of creases formed on the inner surface of a tube after eversion. Our studies of the tube-eversion induced creasing in this paper may trigger more investigations of mechanical instabilities associated with eversion of different structures such as spherical shells.

The work is supported by the National Science Foundation through Grant No. CMMI-1538137. We acknowledge Prof. Robert L. Sah in the department of bioengineering at UCSD for helping us to take the microCT images of an everted tube.

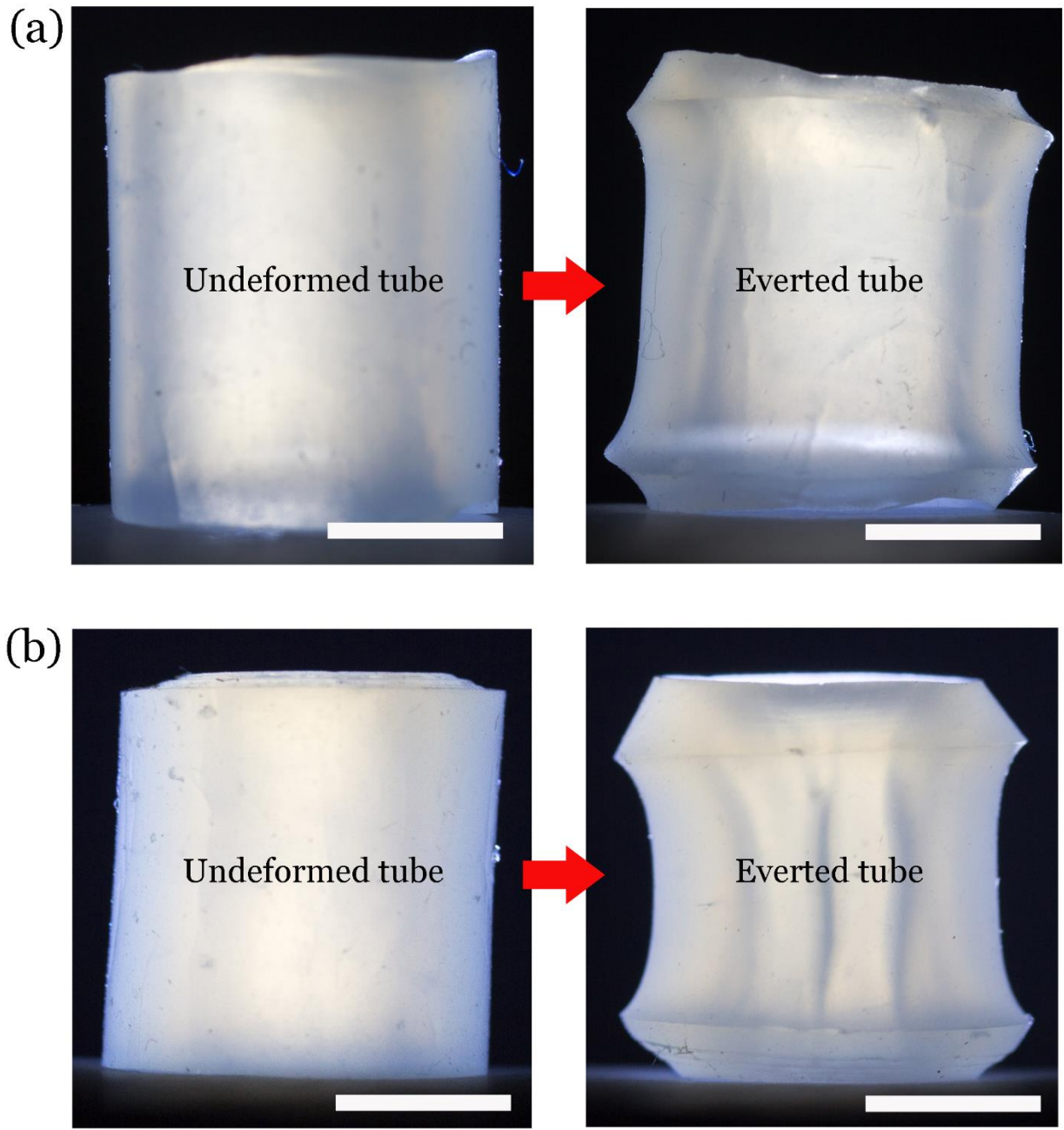

Fig. 1 Crease formation on the inner surface of an everted elastomer tube. (a) If the thickness of the tube is small, after eversion, the inner surface of the tube is smooth; (b) If the thickness of the tube is larger than a critical value, multiple creases form on the inner surface of the everted tube. In the photos above, the outer radius of the tube is 6mm, while thickness of the tube in (a) is 2.5 mm and in (b) is 3.0 mm. The length of the scale bar is 5mm.

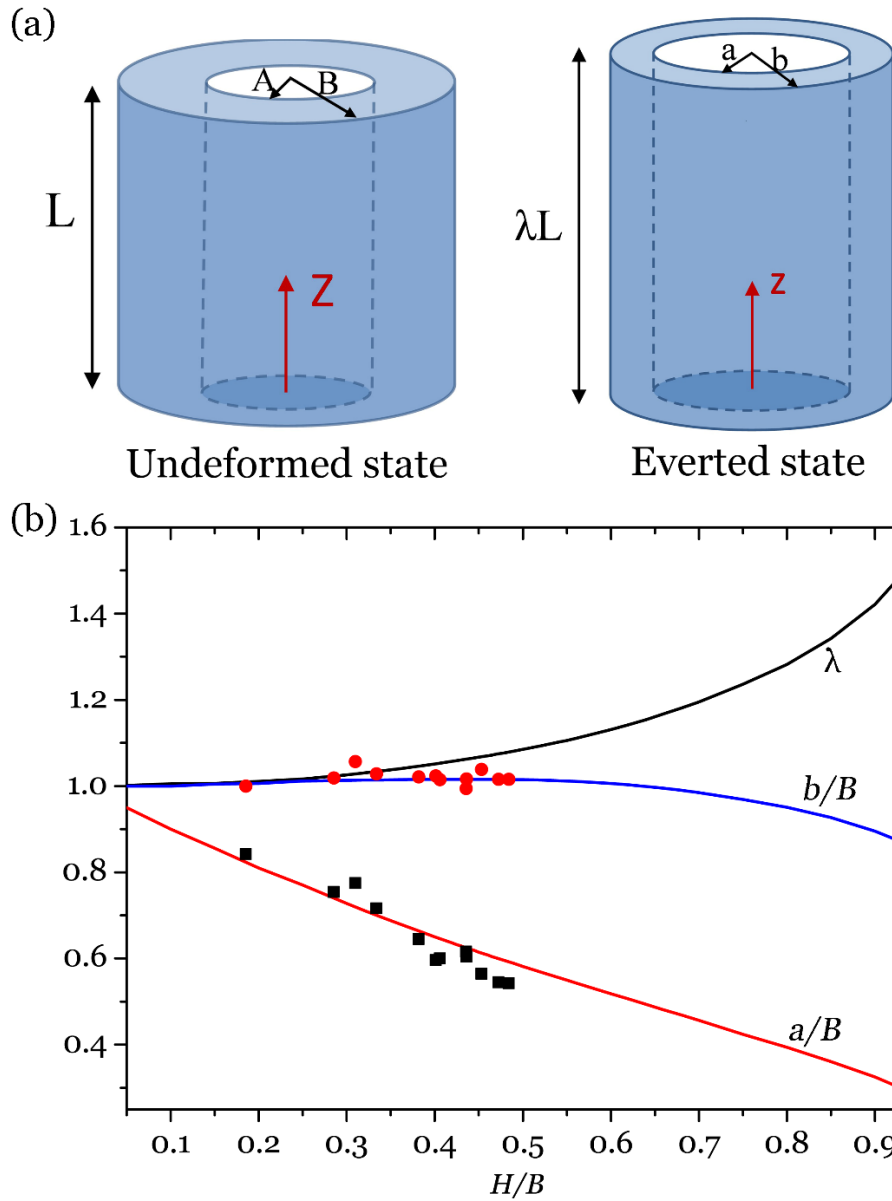

Fig. 2 (a) Schematics of a tube in the undeformed state and everted state. (b) After eversion, the stretch in the axial direction $\lambda$, the inner radius $a$ and the outer radius $b$ of the tube are functions of the tube thickness $H$ of the undeformed state. In the calculation, we ignore all possible mechanical instabilities. The comparisons between our analytical predictions and experimental measurements of the inner radius $a$ (square dots) and outer radius $b$ (circle dots) of everted tubes are shown.

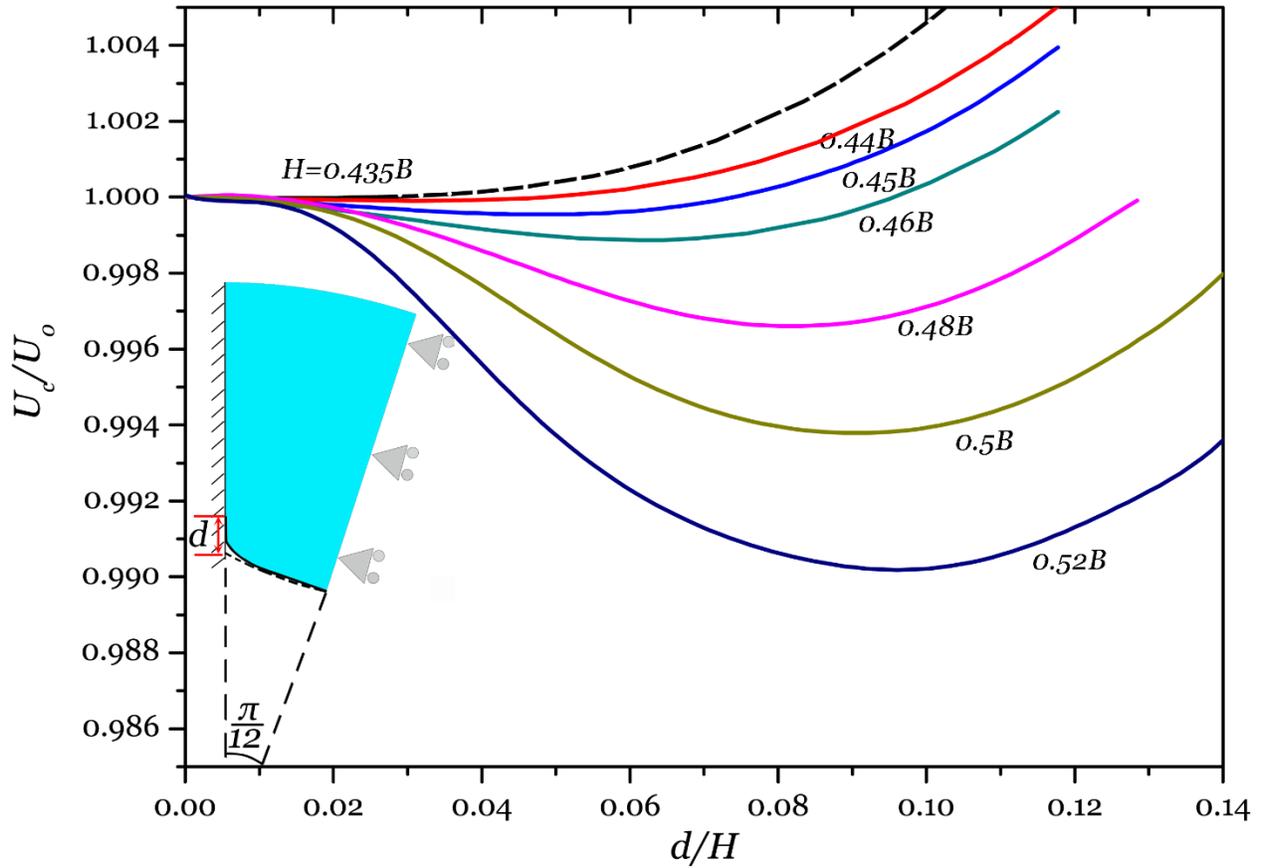

Fig. 3 Normalized strain energy of an everted tube with different thicknesses as a function of crease depth $d$. When the thickness $H$ is smaller than the critical thickness: $0.435B$ (dash line), the strain energy of the everted tube increases monotonically with the increase of crease depth. When the thickness $H$ is larger than the critical thickness, the strain energy of the everted tube has a minimum value for a finite crease depth. In the above calculation, we assume there are 12 creases distributed periodically around the inner surface of the everted tube.

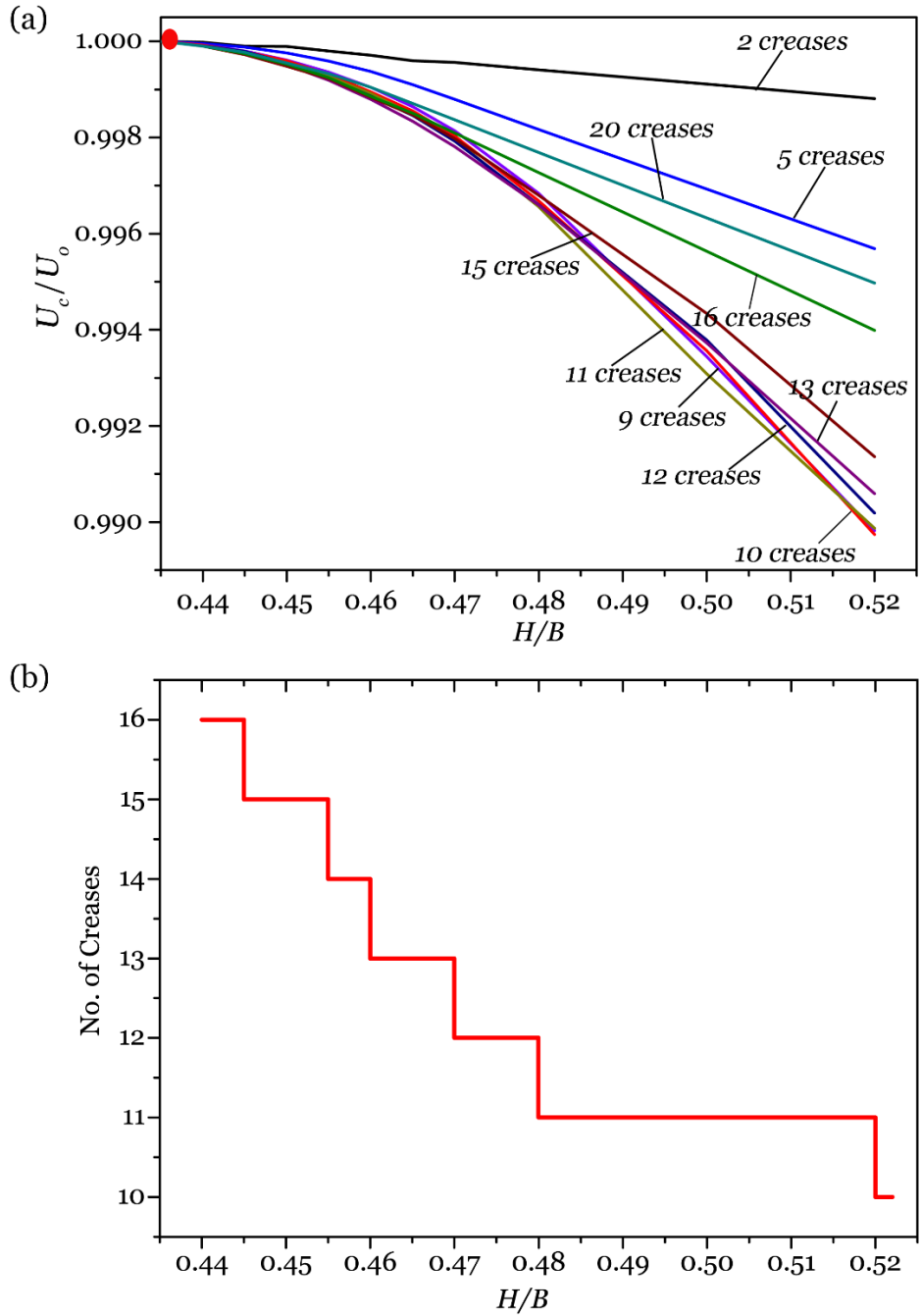

Fig. 4 (a) Normalized strain energy of an everted tube with different number of creases. (b) Number of creases on its inner surface of everted tube with minimal strain energy.

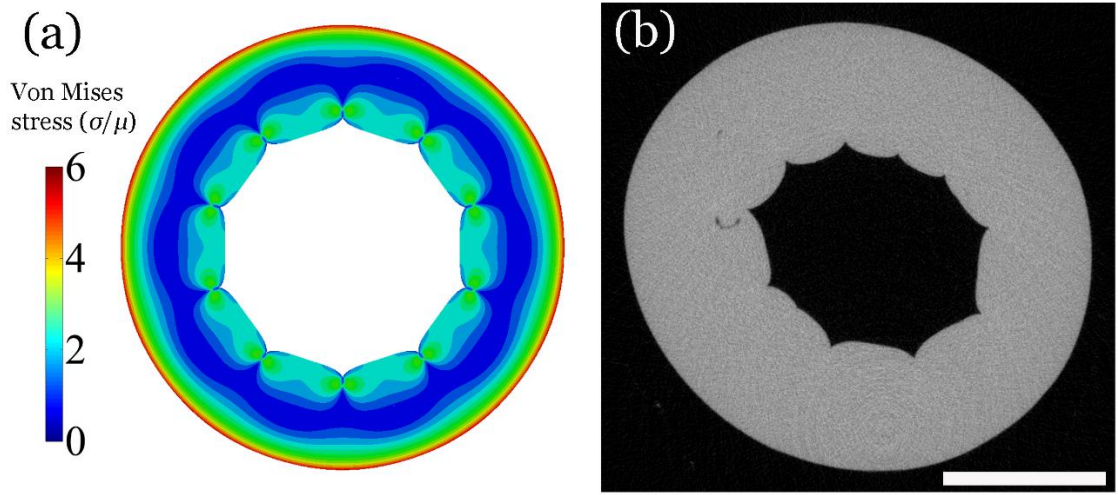

Fig. 5 The comparison between the predicted and microCT image of a cross-section of an everted tube with the initial thickness of $H/B$=0.52. The color in (a) stands for the Von Mises stress normalized by the shear modulus $\mu$ of the elastomer. The length of the scale bar is 5mm.

**References**

1. S. Höhn, A. R. Honerkamp-Smith, P. A. Haas, P. K. Trong, and R. E. Goldstein, Physical review letters **114**, 178101 (2015).
2. K. S. Freeman, G. A. Lewbart, W. P. Robarge, C. A. Harms, J. M. Law, and M. K. Stoskopf, American journal of veterinary research **70**, 1087 (2009).
3. G. A. Lewbart, *Invertebrate medicine* (John Wiley & Sons, 2011).
4. T. Mizuno *et al.*, Journal of Biomedical Materials Research Part B: Applied Biomaterials **102**, 1038 (2014).
5. D. Haughton and A. Orr, International journal of non-linear mechanics **30**, 81 (1995).
6. D. Haughton and A. Orr, International journal of solids and structures **34**, 1893 (1997).
7. M. S. Pour and Y. Fu, SIAM Journal on Applied Mathematics 62, 1856 (2002).
8. O. Varga, *Stress-strain behavior of elastic materials: selected problem of large deformation* (Interscience Publishers, 1966).
9. P. Chadwick and E. Haddon, IMA Journal of Applied Mathematics **10**, 258 (1972)
10. J. Ericksen, Journal of Applied Mathematics and Mechanics/Zeitschrift für Angewandte Mathematik und Mechanik **35**, 382 (1955).
11. S. S. Antman, Archive for Rational Mechanics and Analysis **70**, 113 (1979).
12. R. S. Rivlin, Philosophical Transactions of the Royal Society of London A: Mathematical, Physical and Engineering Sciences **242**, 173 (1949).
13. C. Truesdell, *Rational thermodynamics* (Springer Science & Business Media, 2012).
14. X. Liang and S. Cai, Applied Physics Letters **106**, 041907 (2015).
15. W. Hong, X. Zhao, and Z. Suo, Applied Physics Letters **95**, 111901 (2009).
16. D. Chen, J. Yoon, D. Chandra, A. J. Crosby, and R. C. Hayward, Journal of Polymer Science Part B: Polymer Physics 52, 1441 (2014).
17. L. R. G. Treloar, *The physics of rubber elasticity* (Oxford university press, 1975).
18. Gent and I. Cho, Rubber Chemistry and Technology **72**, 253 (1999).
19. M. Biot, Applied Scientific Research, Section A 12, 168 (1963).

**Appendix A. Axisymmetric deformation in an everted tube**

This section reviews the equilibrium solution of an everted elastomer tube with axisymmetric deformation. As illustrated in Fig. 2a, the inner radius and outer radius of an undeformed cylindrical tube are denoted by *A* and *B*, respectively. The thickness of the tube is denoted by *H=B-A*. After the eversion, except of the region close to the ends of the tube, most part of the tube remains cylindrical shape with inner radius *a* and outer radius *b*. A material particle with distance *R* from the center of an undeformed tube moves to the position with distance *r* from the center of the tube in the everted state. The deformation geometry enables us to calculate the hoop stretch: $\lambda_\theta = r/R$ and radial stretch: $\lambda_r = -dr/dR$ in the tube, respectively.

Following the literature [1], tube eversion can be regarded as a general plane strain problem, with homogenous stretch $\lambda_z = \lambda$ in the axial direction. After eversion, the inner surface of an undeformed tube becomes the outer surface of the tube in the everted state, and the outer surface of an undeformed tube becomes the inner surface of the tube in the everted state, namely,

$$r(A) = b, \quad r(B) = a. \tag{A1}$$

The elastomer is taken to be incompressible, so that we have,

$$B^2 - R^2 = \lambda(r^2 - a^2), \tag{A2}$$

and the deformation field can be further written as,

$$r(R) = \sqrt{(B^2 - R^2)/\lambda + a^2}. \tag{A3}$$

When the axial stretch $\lambda$ and the inner radius *a* in the deformed state are known, the deformation field of the everted tube *r(R)* can be fully determined.

Based on (A3), we can calculate the hoop stretch and radial stretch:

$$\lambda_\theta = \sqrt{\left((B/R)^2 - 1\right)/\lambda + (a/R)^2}, \tag{A4}$$

$$\lambda_r = 1/\sqrt{\left((B/R)^2 - 1\right)\lambda + (a\lambda/R)^2}. \tag{A5}$$

We assume that the elastomer can be described by Neo-Hookean model [2]. The Cauchy stress along the radial, the hoop and the axial directions of the tube can be written as

$$\sigma_r = \mu\lambda_r^2 - p, \tag{A6}$$

$$\sigma_\theta = \mu\lambda_\theta^2 - p, \tag{A7}$$

$$\sigma_z = \mu\lambda_z^2 - p, \tag{A8}$$

where $\mu$ is the small-deformation shear modulus of the elastomer and $p$ is hydrostatic pressure.

Plugging (A6)–(A8) into the stress balance equation (2) in the main text and using the boundary condition $\sigma_r = 0$ at $r=a$, we can obtain,

$$p(r) = \frac{\mu(\lambda a^2 - \lambda r^2 + B^2)}{\lambda^2 r^2} + \frac{\mu(\lambda a^2 + B^2)}{2\lambda^2}\left(\frac{1}{a^2} - \frac{1}{r^2}\right) + \frac{\mu}{\lambda}\ln\frac{aR}{Br}. \tag{A9}$$

The boundary condition $\sigma_r = 0$ at $r=b$ gives that

$$\frac{(\lambda a^2 + B^2)}{2\lambda^2}\left(\frac{1}{a^2} - \frac{1}{b^2}\right) + \frac{1}{\lambda}\ln\frac{aA}{Bb} = 0. \tag{A10}$$

Following Rivlin [3], the relaxed boundary condition of zero resultant force at the end of the tube requires:

$$\int_a^b\left(\frac{\lambda a^2 - \lambda r^2 + B^2}{\lambda^2 r} - \lambda^2 r - \int_a^b\left(\frac{(\lambda a^2 - \lambda r^2 + B^2)(b^2 - r^2)}{2\lambda^2 r^3} - \frac{r(b^2 - r^2)}{2(\lambda a^2 - \lambda r^2 + B^2)}\right)dr\right)dr = 0 \tag{A11}$$

Both axial stretch $\lambda$ and the inner radius $a$ of the inverted tube can be calculated numerically from equations (A10) and (A11). Therefore, with $\lambda$ and $a$ known, the deformation field $r(R)$, the radial stretch $\lambda_r$ and hoop stretch $\lambda_\theta$ can be calculated using (A3), (A4) and (A5).

Fig. 2b plots the axial stretch, the inner radius $a$ and outer radius $b$ of the inverted tube as functions of the thickness of the tube in the undeformed state. In Fig. S1a, we plot the field of radial stretch and hoop stretch in an everted tube for the thickness $H=0.5B$. In Fig. S1b, we plot the corresponding radial stress and hoop stress in the everted tube. In Fig. S2, we plot the radial stretch, hoop stretch and axial stretch of the inner surface of an everted tube with different thicknesses.

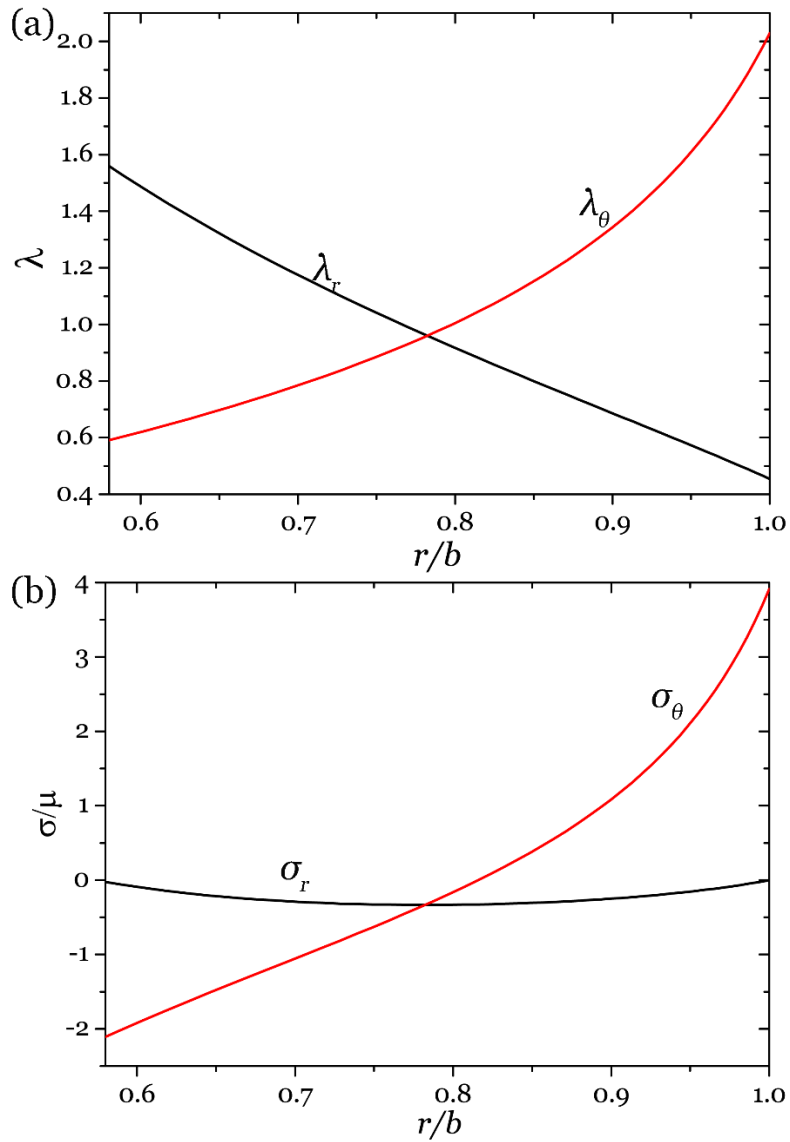

Fig. S1 (a) Distribution of radial stretch and hoop stretch in an everted tube with the thickness *H=0.5B*. (b) Distribution of radial stress and hoop streess in the everted tube with the thickness *H=0.5B*.

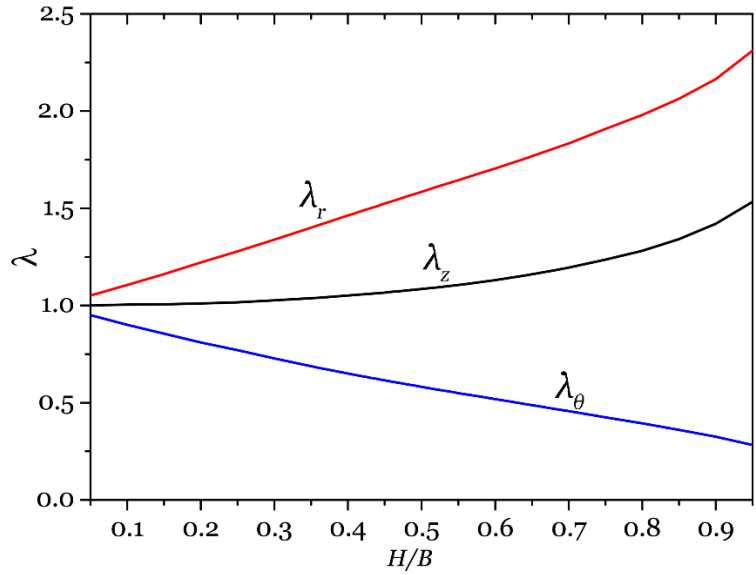

Fig. S2 Radial stretch, hoop stretch and axial stretch of the inner surface of an everted tube as functions of the tube thickness $H$.

**Appendix B. Linear stability analyses of an everted tube**

The section summarizes the linear stability analysis of the everted tube. The stretch and stress field of an everted tube with axisymmetric deformation have been obtained in *Appendix A*. Each material point of a tube in the undeformed state is described by the reference coordinate **X**, which moves to current coordinate **x** after the eversion. The axisymmetric deformation filed of the elastomer is given by **x**⁰(**X**), with deformation gradient defined as,

$$F_{iK}^0 = \frac{\partial x_i^0(\mathbf{X})}{\partial X_K}. \tag{B1}$$

To obtain the critical conditions of wrinkling of the elastomeric tube after eversion, we adopt linear perturbation analysis [4], by perturbing the equilibrium solution $x_i^0(\mathbf{X})$ with a state of infinitesimal displacement $\mathbf{u}(\mathbf{x})$. Using Neo-Hookean material model [2], we can obtain the corresponding perturbation of Cauchy stress as

$$\tilde{\sigma}_{ij} = \mu F_{jK} F_{pK} L_{ip} + p L_{ji} - \tilde{p} \delta_{ij}, \tag{B2}$$

where $\tilde{\sigma}_{ij}$ and $\tilde{p}$ are the perturbed true stress and hydrostatic pressure, and $L_{ij} = \partial u_i / \partial x_j$. The incompressible condition can be expressed as $L_{ii} = 0$.

The perturbation of Cauchy stress also needs to satisfy force balance equation, namely,

$$\frac{\partial \tilde{\sigma}_{ij}}{\partial x_j} = 0. \tag{B3}$$

In general, the displacement perturbation $\mathbf{u}(\mathbf{x})$ can be decomposed into the radial component $u_r(r,\theta)$ and the hoop component $u_\theta(r,\theta)$. Before the perturbation, the deformation field of the tube is axisymmetric and under the generalized plane strain condition. Consequently, the gradient of the displacement perturbation is given by,

$$L_{rr} = \frac{\partial u_r}{\partial r}, L_{\theta\theta} = \frac{u_r}{r} + \frac{1}{r}\frac{\partial u_\theta}{\partial \theta}, L_{r\theta} = \frac{1}{r}\frac{\partial u_r}{\partial \theta} - \frac{u_\theta}{r}, L_{\theta r} = \frac{\partial u_\theta}{\partial r}. \tag{B4}$$

The incompressible condition is,

$$\frac{\partial u_r}{\partial r}+\frac{u_r}{r}+\frac{1}{r}\frac{\partial u_\theta}{\partial \theta}=0. \tag{B5}$$

The perturbed stress $\tilde{\sigma}_{ij}$ in (B2) takes the form,

$$\begin{aligned}\tilde{\sigma}_{rr}&=\left(\mu\lambda_r^2+p\right)L_{rr}-\tilde{p},\\ \tilde{\sigma}_{\theta\theta}&=\left(\mu\lambda_\theta^2+p\right)L_{\theta\theta}-\tilde{p},\\ \tilde{\sigma}_{r\theta}&=\mu\lambda_\theta^2 L_{r\theta}+pL_{\theta r},\\ \tilde{\sigma}_{\theta r}&=\mu\lambda_r^2 L_{\theta r}+pL_{r\theta}.\end{aligned} \tag{B6}$$

The perturbed stress balance equation (B3) can be expressed in the polar coordinate as,

$$\begin{aligned}\frac{\partial \tilde{\sigma}_{rr}}{\partial r}+\frac{1}{r}\frac{\partial \tilde{\sigma}_{r\theta}}{\partial \theta}+\frac{\tilde{\sigma}_{rr}-\tilde{\sigma}_{\theta\theta}}{r}&=0,\\ \frac{\partial \tilde{\sigma}_{\theta r}}{\partial r}+\frac{1}{r}\frac{\partial \tilde{\sigma}_{\theta\theta}}{\partial \theta}+\frac{\tilde{\sigma}_{\theta r}+\tilde{\sigma}_{r\theta}}{r}&=0.\end{aligned} \tag{B7}$$

The boundary condition for the perturbed stress can be expressed by,

$$\tilde{\sigma}_{rr}=0, \tilde{\sigma}_{\theta r}=0. \tag{B8}$$

By setting the perturbed displacement field as

$$\begin{aligned}u_r(r,\theta)&=f(r)\cos(m\theta),\\ u_\theta(r,\theta)&=g(r)\sin(m\theta),\\ \tilde{p}(r,\theta)&=k(r)\cos(m\theta),\end{aligned} \tag{B9}$$

where $f(r)$, $g(r)$ and $k(r)$ are real function and $m$ is the wave number. Substituting (B9) into (B4)-(B7), we obtain that,

$$\begin{aligned}&\frac{\mu R^2}{m^2\lambda^2}f^{IV}+\frac{2\mu(R^2-2\lambda r^2)}{m^2\lambda^2 r}f'''-\left(\frac{p'r}{m^2}+\frac{\mu((4+m^2)R^2+10\lambda r^2)}{m^2\lambda^2 r^2}+\frac{(m^2+1)\mu r^2}{m^2 R^2}\right)f''\\ &-\left(\frac{p''r}{m^2}+\frac{2}{m^2}p'-\frac{\mu((4+m^2)R^2+2(m^2+2)\lambda r^2)}{m^2\lambda^2 r^3}+\frac{3\mu(m^2+1)rR^2+2\mu(m^2+1)\lambda r^3}{m^2 R^4}\right)f'\\ &+\left(\frac{\mu(m^4-1)}{m^2 R^2}+\frac{2\mu(m^2-1)\lambda r^2}{m^2 R^4}+\frac{m^2-1}{m^2}p''\right)f=0,\end{aligned} \tag{B10}$$

where $r=r(R)$ and $p=p(R)$ can be obtained from (A3) and (A9) in *Appendix A*. Substituting (B9) into the boundary condition (B8) yields,

$$\frac{R^2}{\lambda^2 m^2} f''' + \frac{2(R^2 - \lambda r^2)}{\lambda^2 m^2} f'' - \left( \frac{r^2}{R^2} + \frac{2R^2}{\lambda^2 r^2} + \frac{(R^2 + 2\lambda r^2)}{\lambda^2 m^2 r^2} \right) f' - \frac{(m^2 - 1)(R^2 + 2\lambda r^2)}{\lambda^2 m^2 r^3} f = 0, \quad (B11)$$

$$r^2 f'' + rf' + (m^2 - 1)f = 0, \quad (B12)$$

for both $r=a$ and $b$.

The differential equation (B10) accompanied with the boundary (B11) and (B12) consist an eigenvalue problem for the loading parameter $H$. The result can be resolved numerically by compound matrix method [5].

In Fig. S3, we plot the critical thickness $H$ for wrinkling instability with respect to wave number $m$. The smallest tube thickness for the onset of wrinkling instability in the everted tube is defined as critical thickness, which is $H_{cirt}=0.58B$.

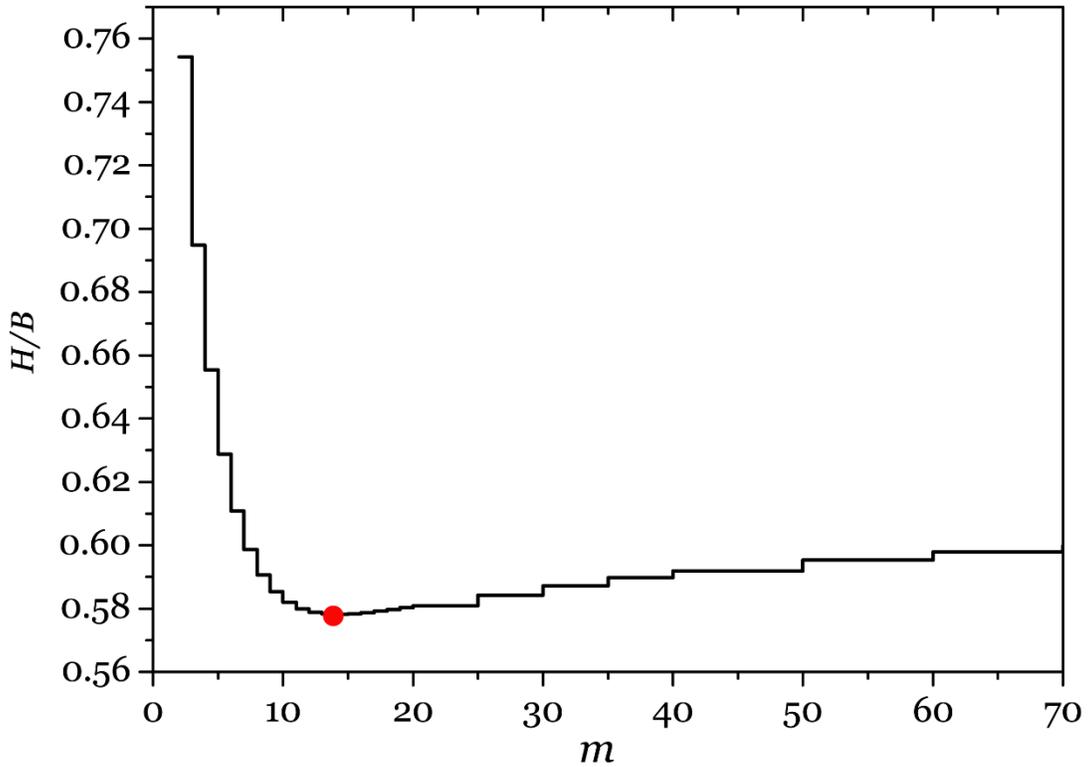

Fig. S3 Critical tube thickness for the wrinkling instability with respect to the wavenumber $m$. The red dot represents the critical thickness for the wrinkling instability in the an everted tube, which is $H_{cirt}=0.58B$.

**Appendix C. Finite element simulation of crease formation in an everted tube**

Creases are localized folds with a singular region of self-contact, around which the strain field is concentrated [6, 7]. Consequently, the critical condition for the onset of creases and the subsequent growth of creases cannot be predicted by linear stability analyses. Following our previous studies [8], in this letter, we conduct finite element simulation using commercial software ABAQUS to predict the critical condition of crease initiation and study the growth of creases in an everted cylindrical tube.

To avoid simulating the complex finite deformation of tube eversion process, the equilibrium stress state after eversion is introduced as initial stress through the user subroutine SIGINI in ABAQUS [8]. The stress field is obtained from the equilibrium analysis in *Appendix A* and the subroutine is called at the start of the crease analysis.

We assume that the multiple creases periodically distribute in the inner surface of an everted tube. Therefore, only half crease with symmetric boundary condition is adopted in the simulation as shown in the inset of Fig. 3, with $\theta$ being the sector angle. The number of creases is determined by $\pi/\theta$. To calculate the strain energy of the everted tube with creases, we apply a radial displacement $d$ at a point in the inner surface of the everted tube to induce the formation of a crease. The inner and outer surfaces of the tube are set to be traction free.

In the finite element simulation, the plane strain hybrid element CPE6MH is adopted. To resolve stress/strain field around the crease tip, the mesh size in our simulation is set to be much smaller than the crease depth $d$. In addition, mesh-to-mesh solution mapping [8] is adopted in our simulation when the elements deform significantly from their original configurations and become severely distorted during the crease formation. The old, deformed mesh is replaced by a new mesh of better quality. The solutions are mapped from the old mesh to the new mesh so that the analysis can continue.

# Reference

[1]. D. Haughton and A. Orr, International journal of non-linear mechanics **30**, 81 (1995).

[2]. L. R. G. Treloar, *The physics of rubber elasticity* (Oxford university press, 1975).

[3]. R. S. Rivlin, Philosophical Transactions of the Royal Society of London A: Mathematical, Physical and Engineering Sciences 242, 173 (1949).

[4]. M. Biot, Applied Scientific Research, Section A 12, 168 (1963).

[5]. B. Ng and W. Reid, Journal of Computational Physics **58**, 209 (1985).

[6]. S. Cai, D. Chen, Z. Suo, and R. C. Hayward, Soft Matter **8**, 1301 (2012).

[7]. W. Hong, X. Zhao, and Z. Suo, Applied Physics Letters **95**, 111901 (2009).

[8]. X. Liang and S. Cai, Applied Physics Letters 106, 041907 (2015).

[9]. ABAQUS. 2008 ABAQUS analysis user's manual, version 6.8. See http://www.simula.com.